\documentclass[aps,twocolumn,groupedaddress,amssymb]{revtex4}

\usepackage{latexsym}
\usepackage{amsbsy}
\usepackage{amsmath}
\usepackage{graphicx}
\usepackage{xcolor}
\usepackage{hyperref} 
\usepackage[capitalise]{cleveref} 
\hypersetup{
    colorlinks=true,
    linkcolor=black,
    urlcolor=blue,
    citecolor=blue
}
\usepackage{nameref}  

\begin{document}

\author{Shreyashi Sinha$^1$, Ayan Jana$^2$, Suchanda Mondal$^3$, Ravi Prakash Singh$^3$, Manoranjan Kumar$^2$ and Sujit Manna$^1$\footnote{Contact author: smanna@physics.iitd.ac.in}}
\affiliation{$^1$Department of Physics, {Indian Institute of Technology Delhi}, Hauz Khas, New Delhi 110016, India}
\affiliation{$^2$Department of Condensed Matter Physics, { S. N. Bose National Centre for Basic Sciences}, Kolkata, 7000106, India}
\affiliation{$^3$Department of Physics, {Indian Institute of Science Education and Research Bhopal}, Bhauri, Madhya Pradesh 462066, India}

\preprint{APS/123-QED}

\title{Nanoscale Electronic Phase Separation Driven by Fe-site Ordering in Fe\textsubscript{5-x}GeTe\textsubscript{2}}


\date{\today}

\begin{abstract}

Understanding how local structural order governs electronic correlations is essential for revealing the microscopic mechanism underlying emergent behavior in two-dimensional magnets. In the layered van der Waals ferromagnet Fe\textsubscript{5–x}GeTe\textsubscript{2}, intrinsic Fe-site disorder provides a natural platform to probe this interplay. Here, we establish a direct atomic-scale correlation between Fe-site ordering and local electronic structure by combining high-resolution scanning tunneling microscopy with density functional theory calculations. Scanning tunneling microscopy resolves two coexisting surface phases: a $\sqrt{3} \times \sqrt{3}$ superstructure associated with ordered Fe(1) configurations, and an undistorted $1 \times 1$ hexagonal Te lattice in Fe(1)-deficient regions. Spatially resolved spectroscopy shows that the $\sqrt{3}$-ordered domains exhibit metallic behavior, whereas Fe(1) vacant areas display a suppressed density of states(DOS) near the Fermi level, indicative of pseudogapped electronic states. The nanoscale coexistence of these distinct electronic responses provides direct evidence of electronic phase separation driven by Fe-site ordering. First-principles calculations reveal that symmetry-allowed hybridization between Fe 3d and Te 5p orbitals reconstructs the low-energy electronic structure, giving rise to the contrasting tunneling signatures of ordered and disordered phases. Bias-dependent local DOS simulations reproduce the experimentally observed contrast evolution and reveal that hybridization-induced out-of-plane orbital character governs the spatial modulation of tunneling conductance. These results provide a microscopic framework linking atomic-scale structural order to nanoscale electronic inhomogeneity in van der Waals magnets.

\end{abstract}

\maketitle

\section{Introduction} 

Magnetic van der Waals (vdW) materials have emerged as a versatile platform for exploring correlated phenomena in low dimensions, owing to their atomically thin structure and weak inter-layer coupling \cite{huang2017layer, gong2017discovery}. In these systems, magnetic exchange interactions are largly confined within individual layers, while the quasi two-dimensional(2D) lattice supports strong coupling among spin, charge, and orbital degrees of freedom \cite{gong2019two}. This structural flexibility gives rise to a broad spectrum of ground states, including metallic, semiconducting and insulating phases, as well as ferro- and antiferromagnetic order. Such tunability has positioned vdW magnets at the forefront of research on emergent electronic phases, interfacial engineering, and spintronic functionality \cite{gibertini2019magnetic, burch2018magnetism, huang2020emergent,gong2019two}. Despite rapid progress, however, beyond enhancing the Curie temperature (T\textsubscript{c}), a central unresolved issue concerns how intrinsic structural disorder, ubiquitous in many vdW magnets modulates local electronic correlations and potentially drives nanoscale electronic inhomogeneity. Establishing a direct microscopic link between atomic-scale structural motifs and electronic phase behavior remains a critical challenge \cite{huang2020emergent}.

Among the high T\textsubscript{c} vdW ferromagnets, Fe\textsubscript{n}GeTe\textsubscript{2} (n = 3, 4, 5) series has emerged as particularly promising platform \cite{may2016magnetic, chowdhury2021unconventional, li2018patterning}. Fe\textsubscript{3}GeTe\textsubscript{2} and Fe\textsubscript{4}GeTe\textsubscript{2} exhibits Curie temperature of $\sim$220 K \cite{may2016magnetic} and $\sim$270 K \cite{seo2020nearly} respectively, while, Fe\textsubscript{5–x}GeTe\textsubscript{2} achieves the highest T\textsubscript{c}, ranging from $\sim$260–330 K in bulk crystals and even exceeding room temperature in thin flakes \cite{zhang2020itinerant, may2019physical, chen2022revealing}. The enhanced T\textsubscript{c} in Fe\textsubscript{5–x}GeTe\textsubscript{2} is attributed to strengthened Fe–Fe exchange interactions arising from increased Fe site occupancy \cite{he2024spin, stahl2018van,li2020magnetic}. Beyond its magnetic robustness, this compound displays an unusual degree of structural complexity arising from split-site occupancies of Fe(1) and Ge atoms \cite{ly2021direct, ershadrad2022unusual}, along with the presence of Fe-site vacancies \cite{luo2025direct}. These structural features introduce nanoscale disorder and local symmetry breaking, which strongly influence both the magnetic exchange pathways and the itinerant charge carriers. As a result, the delicate interplay between ordering on the Fe(1) sublattice, vacancy formation, and electronic correlations gives rise to a rich variety of electronic ground states \cite{walve2024unveiling}. Yet, most investigations have relied on bulk-sensitive probes, leaving the spatially resolved impact of Fe-site ordering and vacancy distribution on the electronic structure largely unexplored. The central motivation of this study is to investigate how these local structural motifs couple to the electronic properties of Fe\textsubscript{5–x}GeTe\textsubscript{2}.

\begin{figure*}
   \centering
       \includegraphics[width=\textwidth]{"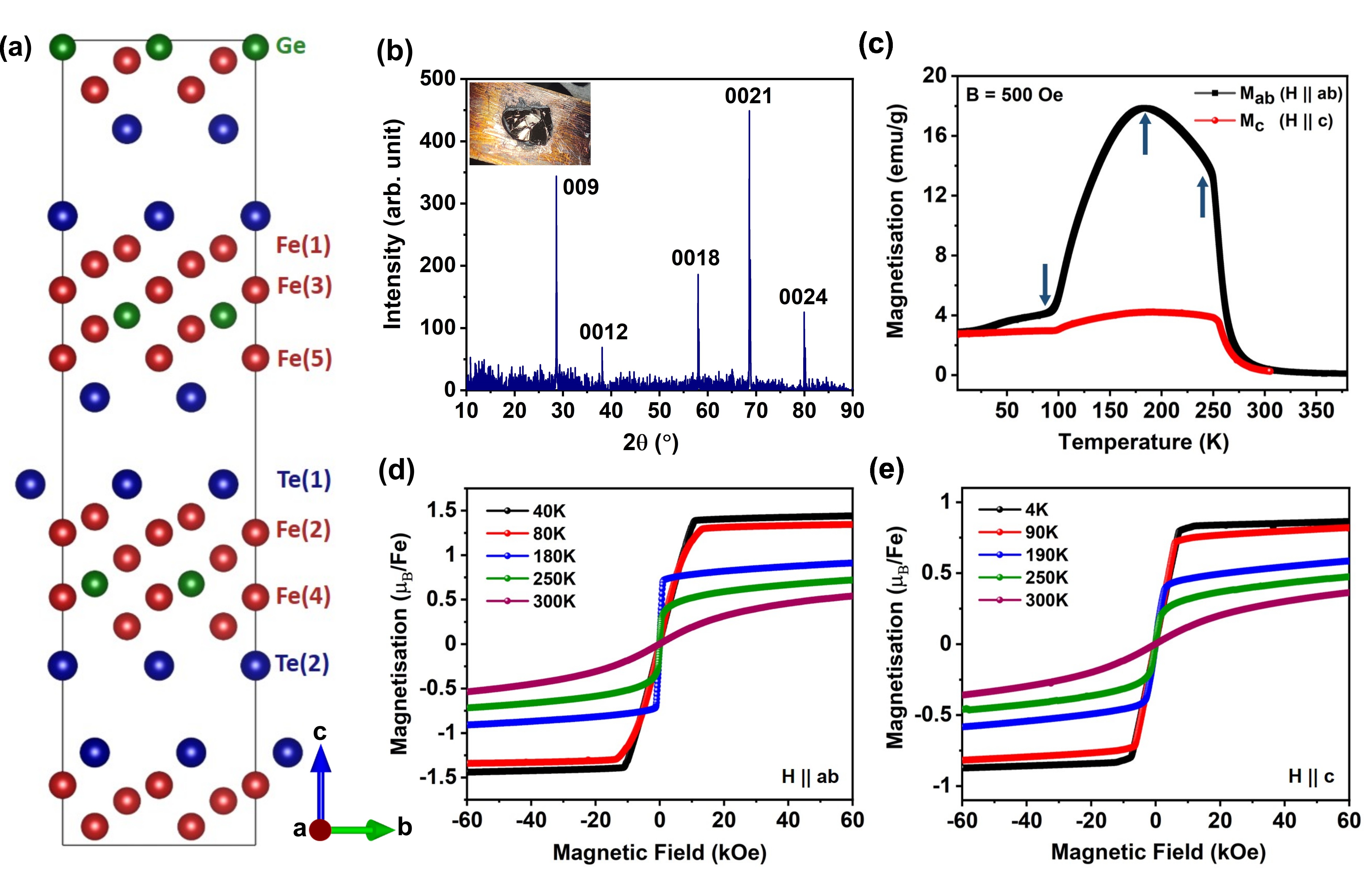"}
    \caption{(a) Side view of the crystal structure of Fe\textsubscript{5–x}GeTe\textsubscript{2}. The rectangle outlines the unit cell, and the positions of Fe, Ge, and Te atoms at their respective sites are indicated. (b) Representative x-ray diffraction pattern collected from a crystal facet with the c-axis oriented normal to the surface, (inset) an optical image of the corresponding single crystal. (c–e) Magnetic characterization of bulk Fe\textsubscript{5–x}GeTe\textsubscript{2} single crystals. (c) Temperature dependence of magnetization under zero-field cooling for in-plane ($H \parallel ab$, M\textsubscript{ab}) and out-of-plane ($H \parallel c$, M\textsubscript{c}) applied fields of 500 Oe. The blue arrows mark transition temperatures 90K, 180K and 250K. (d) M–H curves measured with $H \parallel ab$ at five temperatures from 40K to room temperature, including the transition temperatures. (e) M–H curves measured with $H \parallel c$ at five temperatures from 4K to 300K.}
    \label{Figure1}
\end{figure*}

In this work, we aim to understand the role of Fe(1) ordering in stabilizing different superstructures, such as the $\sqrt{3} \times \sqrt{3}$ phase, which break inversion symmetry and can strongly influence electron correlations and magnetic interactions \cite{ly2021direct, lam2025thermal}. Additionally, we examine the impact of Fe(1) vacancies on the local atomic arrangement, as vacancy-induced disorder has been proposed to play a decisive role in determining electronic transport in vdW magnets. By correlating scanning tunneling microscopy (STM) observations with spectroscopic measurements, we seek to establish a direct link between structural motifs and the emergence of distinct electronic phases. Finally, this work addresses the broader question of how local structural order can stabilize metallic versus pseudogapped ground states within the same material family. Our results demonstrate that Fe\textsubscript{5–x}GeTe\textsubscript{2} exhibits a striking structural-electronic correlation: the $\sqrt{3}a \times \sqrt{3}a$ ordered phase displays metallic characteristics, whereas the Fe(1) vacancy-dominated, undisordered 1a $\times$ 1a phase manifests a pseudogapped electronic state characterized by a pronounced suppression of local density of states near the Fermi level. These findings highlight the crucial role of local Fe(1) configurations in determining macroscopic electronic properties, provides a microscopic foundation insights for understanding disorder driven electronic inhomogeneties in Fe-based vdW systems.

\section{Experimental Details}

Single crystals of  Fe\textsubscript{5–x}GeTe\textsubscript{2} were grown by the chemical vapor transport (CVT) method with I$_2$ as the transport agent. High-purity Fe (5N), Ge (5N), and Te (5N) powders were sealed in an evacuated quartz tube and placed in a two-zone furnace. Crystal growth was carried out under a temperature gradient of 750/700~$^\circ$C for one week. Shiny silver-colored plate-like single crystals with typical lateral dimensions of 3 $\times$ 2 mm$^2$ were obtained and could be readily cleaved along the $ab$-plane \cite{chowdhury2021unconventional}. Magnetic characterization of the sample was performed using a vibrating sample magnetometer (VSM) probe in Physical Property Measurement System (PPMS) manufactured by Cryogenic Limited. Prior to STM measurement, the sample was \textit{in situ} cleaved within the exchange chamber at a base pressure of $2 \times 10^{-10}$ mbar. For this purpose, the sample was first affixed to an STM holder equipped with a cylindrical post, transferred through the load-lock into the ultra-high-vacuum (UHV) system, and subsequently cleaved by applying torque to the post using a wobble stick. Scanning tunneling microscopy (STM) and spectroscopy (STS) measurements were performed at 78 K in the STM chamber of UHV STM USM1300 (Unisoku)\cite{firdosh2025exchange}. STM tips were fabricated from mechanically polished PtIr wire. Differential conductance (dI/dU) spectra were acquired using a lock-in technique, in which an AC modulation ($f = 3.6$ kHz, $V_\text{rms} = 12$–20 mV) was superimposed on the applied bias during ramping. STM images were analyzed and processed with MATLAB and WSxM software \cite{sinha2026imaging,he2018exchange}.

\begin{figure*}
   \centering
       \includegraphics[width=\textwidth]{"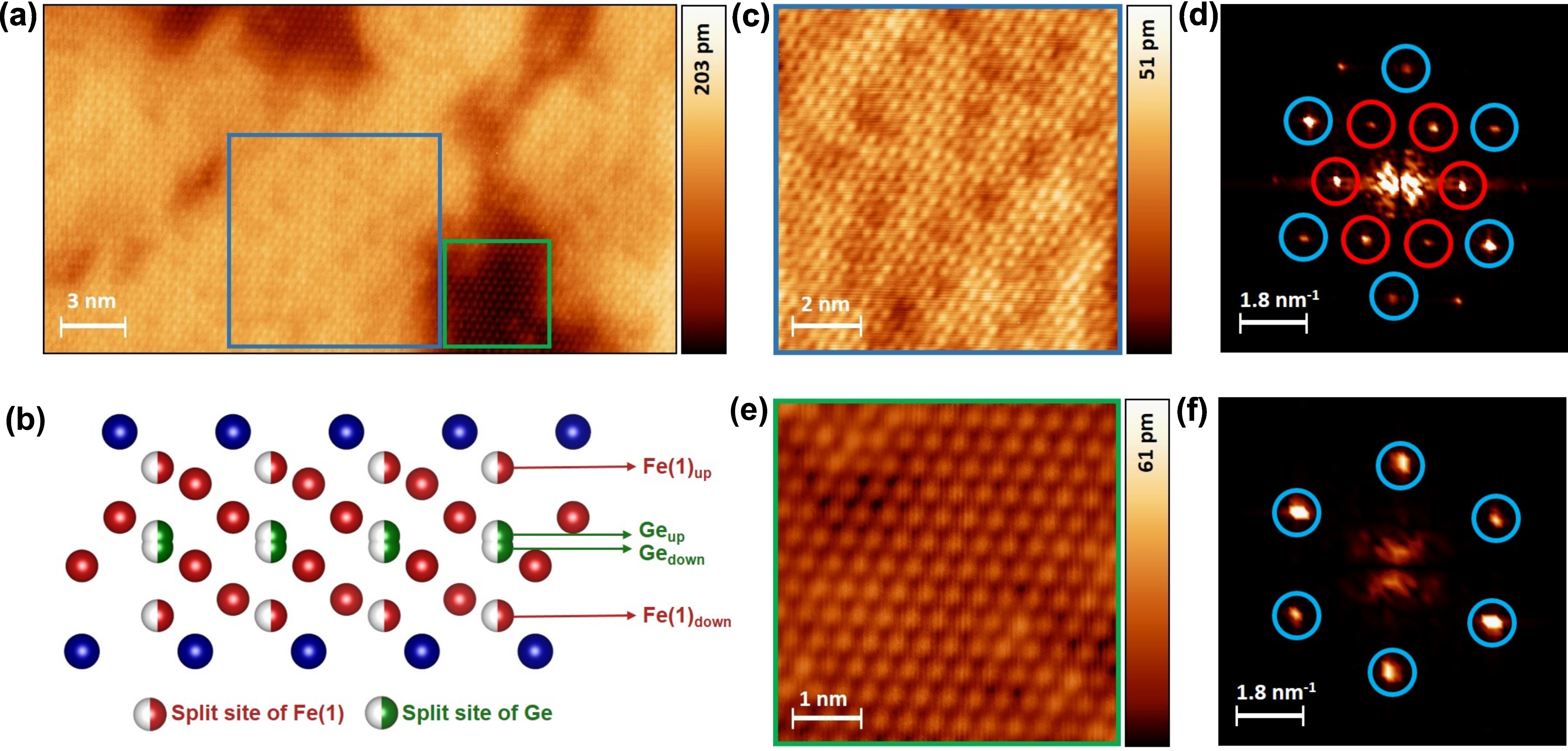"}
    \caption{(a) STM topograph (U = -200 mV, I = 190 pA) of the Te-terminated surface of  Fe\textsubscript{5–x}GeTe\textsubscript{2}, highlighting two regions exhibiting distinct structural orderings.  (b) Crystal structure of Fe\textsubscript{5–x}GeTe\textsubscript{2} portraying possible split-site occupancies of Fe(1) and Ge atoms. (c) STM topograph (10 nm $\times$ 10 nm) (U = -200 mV, I = 190 pA) acquired  from region I (marked by blue square) in panel (a). (d) Corresponding fast Fourier transform (FFT) image, the peaks marked by blue and red circles represent the 1 × 1 hexagonal lattice and the $\sqrt{3} \times \sqrt{3}$ superstructure, respectively. (e) High-resolution (U = -200 mV, I = 190 pA) STM scan (5 nm $\times$ 5 nm) acquired from region II (marked by green square) in panel (a) and (f) its corresponding FFT image showing only the hexagonal lattice.} 
    \label{Figure2}
\end{figure*}

\section{Theoretical Details}
Density Functional Theory (DFT) calculations of the structural and electronic properties were performed using the Vienna Ab initio Simulation Package (VASP)\cite{PhysRevB.54.11169} and  the Perdew, Burke, and Ernzerhof (PBE) functional is used in the framework of generalized gradient approximation (GGA)\cite{PhysRevB.45.13244,PhysRevLett.77.3865},  including spin–orbit coupling (SOC). The experiment reveals two distinct structural phases, and the corresponding surface states are probed using STM. Therefore, we calculate the two-dimensional slab structure, which can be mimicked by placing a vacuum spacing of more than 15~\AA~along the Z-axis to minimize interactions between periodic images. To compare the experimental results for the two structural phases and their corresponding electronic properties, we performed separate density functional theory (DFT) calculations for the $\sqrt{3}a \times \sqrt{3}a$ and $1a \times 1a$ structural configurations. Brillouin-zone integrations were carried out using $\Gamma$-centered Monkhorst--Pack meshes of $5 \times 5 \times 1$ and $9 \times 9 \times 1$ for the $\sqrt{3}a \times \sqrt{3}a$ and $1a \times 1a$ structures, respectively~\cite{PhysRevB.13.5188}.
 The plane-wave cutoff energy was set to 600 eV, and ionic relaxation was carried out until the forces were below 0.0002 eV/~\AA, with an electronic convergence criterion of $1\times10^{-8}$eV. Ionic relaxation was performed using the conjugate gradient method.

The real-space distributions of the energy-integrated local density of states (LDOS) were obtained by integrating the LDOS derived from the Kohn--Sham eigenstates over an energy window extending from the Fermi level $E_{\mathrm{F}}$, to the applied bias voltage. The resulting spatial LDOS maps were visualized using the \textsc{p4vasp}~\cite{p4vasp} and \textsc{VESTA}~\cite{Momma2011} software packages.

\section{Results and Discussion}

Fe\textsubscript{5-x}GeTe\textsubscript{2} is a member of the Fe\textsubscript{n}GeTe\textsubscript{2} family (n = 3, 4, 5), in which each structural unit is composed of a Fe\textsubscript{n}Ge slab encapsulated between two Te layers \cite{ly2021direct}. The structural model, shown in Figure \ref{Figure1}a, depicts the specific lattice positions of Fe, Ge, and Te atoms. Fe\textsubscript{5-x}GeTe\textsubscript{2} crystallizes in a rhombohedral lattice with a centrosymmetric space group $R\overline{3}m$ \cite{may2019ferromagnetism, alahmed2021magnetism, wu2021direct}. Figure \ref{Figure1}b presents the x-ray diffraction pattern obtained from an as-grown facet of a Fe\textsubscript{5-x}GeTe\textsubscript{2} crystal. The presence of only 00l reflections confirms that the vapor-grown single crystals are oriented with the c-axis perpendicular to the surface and the sharp peaks indicate high crystallinity of the sample.

Understanding the magnetic properties of Fe\textsubscript{5-x}GeTe\textsubscript{2} is crucial to elucidate the relationship between its structural ordering and electronic phases, as well as for potential applications in spintronic and magnetic devices. The magnetic behavior of single crystals was investigated through temperature- and field-dependent magnetization measurements. Figure \ref{Figure1}c shows the M-T curves recorded under a small applied magnetic field of 500 Oe along both the in-plane ($H \parallel ab$) and out-of-plane ($H \parallel c$) directions. These are measured upon heating from 2 K to 380 K after zero-field cooling (ZFC). The Curie temperature is estimated to be T\textsubscript{c} = 286 K by fitting the inverse susceptibility ($\chi^{-1}$) vs temperature curve for high-temperature region \cite{he2024spin} for $H \parallel ab$ (see Supporting Information \cite{SI}). The induced moment is larger for in-plane fields than for those applied along the c-axis, reflecting the easy-plane magnetic anisotropy \cite{alahmed2021magnetism} of the system arising from spin–orbit coupling and the layered crystal structure. The broad cusp observed in the temperature range of 90–280 K and additional anomalies at lower temperatures (the transition temperatures marked by blue arrows), can be ascribed to complex spin reorientation or domain rearrangements of the sample \cite{may2019physical, gao2020spontaneous}. To probe these features in detail, isothermal magnetization measurements were carried out at temperatures corresponding to the transition temperatures observed in the M-T curves. The M-H data in Figures \ref{Figure1}d and \ref{Figure1}e  demonstrate strong magnetic anisotropy, with higher saturation fields and distinct coercivity along the hard axis (c-axis) compared to the easy plane (ab-plane). This difference is prominently observed in the temperature range of 90 K-250 K, similar to Figure \ref{Figure1}c. As the temperature increases, the saturating magnetization demonstrates a gradual decrease for both the cases. These measurements highlight the interplay between crystallographic orientation and magnetic response, reflecting the influence of Fe(1) layer ordering and vacancy distribution on the magnetic ground state.

A defining structural characteristic of Fe\textsubscript{5}GeTe\textsubscript{2}, distinguishing it from other Fe\textsubscript{n}GeTe\textsubscript{2} compounds, lies in the possibility of split-site occupancy for both the Fe(1) atoms in the outermost Fe\textsubscript{5}Ge sublayer and the Ge atoms, thereby introducing substantial structural complexity \cite{ly2021direct}. As illustrated in the Figure \ref{Figure2}b, Fe(1) can reside in two alternative positions: Fe(1)\textsubscript{up} (U), located adjacent to the upper Te layer, or Fe(1)\textsubscript{down} (D), positioned near the lower Te layer \cite{lam2025thermal}. The occupation of Fe(1) in either of these equally probable sites induces a corresponding displacement of the Ge atom, resulting in its split position. Specifically, when Fe(1) occupies the upper site above Ge, the Ge atom is shifted downward along the c-axis, and vice versa, ensuring the preservation of the required Fe–Ge bond length \cite{ershadrad2022unusual, ly2021direct}. These orderings are not limited to single Fe(1) atom dynamically changing its location, it occurs within a larger repeating unit, commonly referred to as $\sqrt{3} \times \sqrt{3}$ supercell. Importantly, this ordered configuration of the Fe(1) layer breaks the inversion symmetry of the lattice, giving rise to key magnetic phenomena such as the Dzyaloshinskii–Moriya interaction \cite{ly2021direct}.

In our study, we have identified two distinct types of structural ordering in the Fe\textsubscript{5-x}GeTe\textsubscript{2} sample from the STM topography of the examined regions is presented in Figure \ref{Figure2}a. The first type of ordering, corresponds to a $\sqrt{3}a \times \sqrt{3}a$ periodicity (Figure \ref{Figure2}c), where an additional superstructure is observed on the hexagonal lattice of the top Te layer. The fast Fourier transformation (FFT) of this region displayed in Figure \ref{Figure2}d confirms the presence of a $\sqrt{3} \times \sqrt{3}$ superstructure (highlighted by red circles) relative to the fundamental hexagonal lattice (blue circles). By contrast, the magnified topograph in Figure \ref{Figure2}e, taken from the darker area of Figure \ref{Figure2}a, along with its corresponding FFT image (Figure \ref{Figure2}f), reveals an undistorted hexagonal arrangement of the top Te layer. This undistorted lattice is attributed to local Fe(1) vacancies \cite{ly2021direct, lam2025thermal}, the ordering of which drives the formation of $\sqrt{3} \times \sqrt{3}$ superstructure. These Fe vacancies are frequently reported in Fe\textsubscript{n}GeTe\textsubscript{2} systems and Fe(1) has the maximum tendency to host a vacancy \cite{may2019physical, nguyen2018visualization, ershadrad2022unusual, ly2021direct}. Notably, no domain boundaries were observed between the two regions, indicating that the crystal axes remain well aligned across them. The coexistence of ordered and vacancy-induced undistorted regions suggests strong local variations in Fe occupancy, which can significantly influence the electronic and magnetic properties.

\begin{figure*}
   \centering
       \includegraphics[width=\textwidth]{"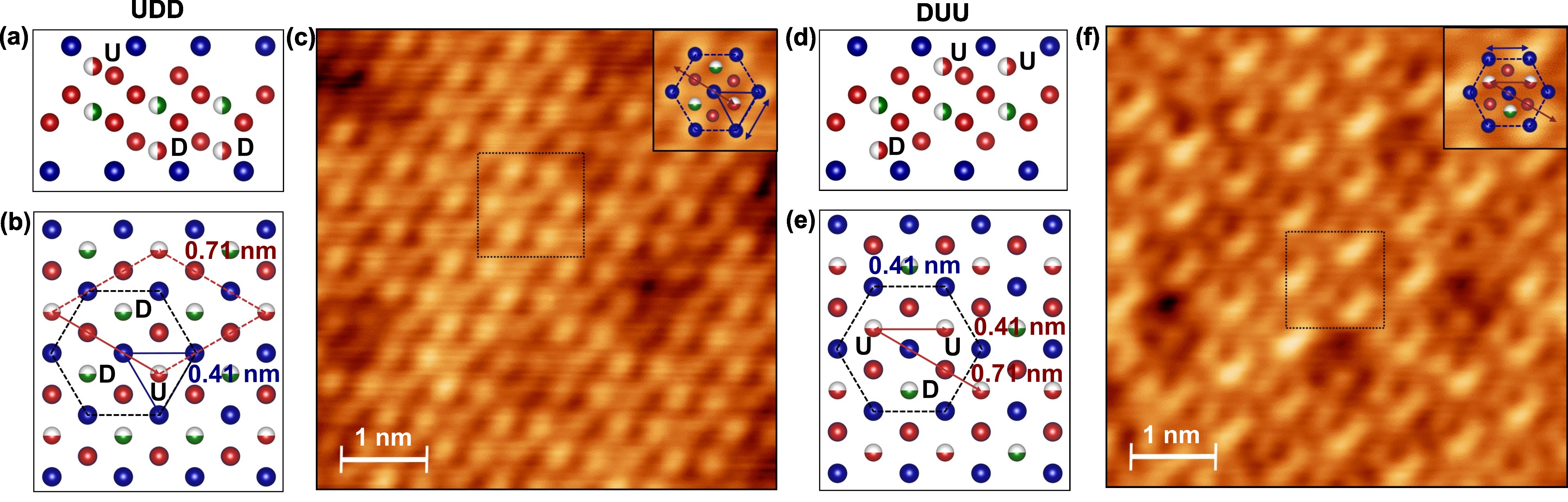"}
    \caption{Schematic representation and STM topographs of the two $\sqrt{3} \times \sqrt{3}$ superstructure phases. (a) Side view and (b) top view atomic model of the UDD configuration. The red rhombus marks the unit cell, while the blue triangle highlights Te trimers. (c) STM topograph (5 nm × 5 nm) of the UDD phase (U = -200 mV, I = 250 pA), (inset) atoms within the hexagonal lattice superimposed for clarity. (d) Side view and (e) top view atomic model of the DUU configuration. (f) STM topograph (5 nm × 5 nm) of the DUU phase (U = -200 mV, I = 250 pA). Half-red, full-red, blue, and half-grey spheres denote Fe(1), Fe, Te, and Ge atoms, respectively.}
    \label{Figure3}
\end{figure*}

\begin{figure}[t]
   \centering
       \includegraphics[width=\columnwidth]{"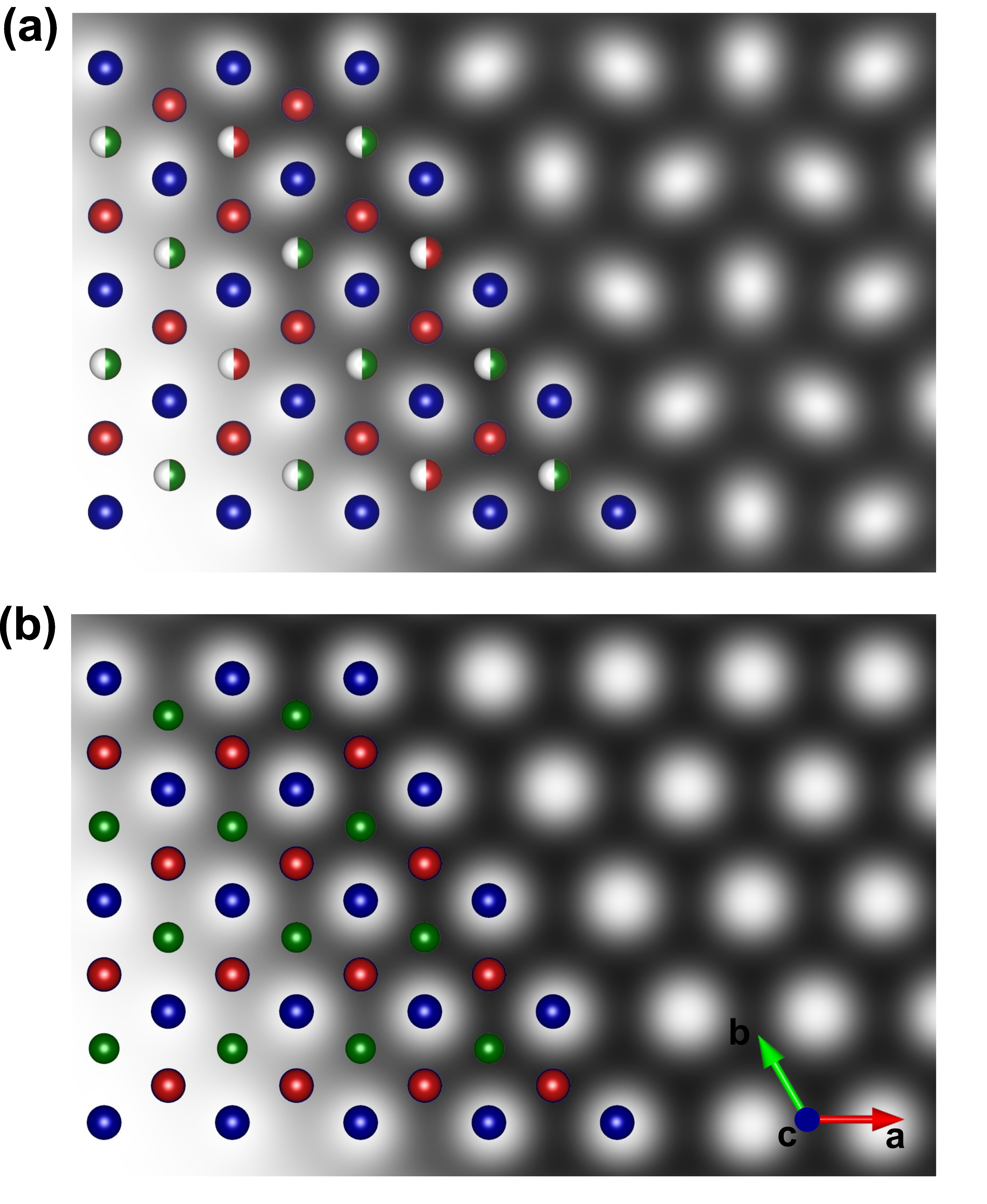"}
    \caption{Simulated STM images obtained from density functional theory calculations, plotted at a fixed isosurface value and overlaid with the corresponding ball model to indicate the atomic positions. (a) Relaxed UDD structure and (b) relaxed undistorted hexagonal structure, shown as top views along the $z$ direction. Bright regions correspond to enhanced local density of states (LDOS) at the selected bias, while darker regions indicate reduced LDOS. Blue, red, and green spheres denote Te, Fe, and Ge atoms, respectively. The crystallographic axes are indicated in panel (b).}
    \label{Figure4}
\end{figure}

Interestingly, two distinct phases of the $\sqrt{3}a \times \sqrt{3}a$ superstructure were observed, which can be directly correlated with the symmetry-dictated orderings of the Fe(1) sublayer. Depending on the relative arrangement of Fe(1) atoms within the reconstructed $\sqrt{3}a \times \sqrt{3}a$ supercell, several configurations are possible, such as Fe(1)\textsubscript{up}–Fe(1)\textsubscript{down}-Fe(1)\textsubscript{down} (UDD), Fe(1)\textsubscript{down}–Fe(1)\textsubscript{up}–Fe(1)\textsubscript{up} (DUU), UDU, and UUU/DDD. Among these, the UDU ordering has been reported as the most favorable due to its lower formation energy \cite{ershadrad2022unusual}, whereas the UUU/DDD states are energetically less stable and therefore rarely observed in Fe\textsubscript{5–x}GeTe\textsubscript{2}. In contrast, UDD/DUU configurations are favored when crystals are quenched from temperatures above a structural transition reported in the literature at T\textsubscript{HT} = 550 K \cite{wu2024reversible}. Figures \ref{Figure3}c and \ref{Figure3}f portray the STM surface morphology of $\sqrt{3}a \times \sqrt{3}a$ superstructures corresponding to the UDD and DUU orderings, respectively, along with schematic atomic models. A clear difference in the atomic arrangement is observed between these two STM topographs: in the UDD phase, two darker spots and one relatively brighter spot are visible among the three Fe(1) sites within the hexagon of six Te atoms, whereas in the DUU phase, one darker spot and two brighter spots are observed. The in-plane lattice constant $a$ was determined to be approximately 0.41 nm \cite{wu2021direct} for both phases, as obtained from line profiles drawn across the bright Te atoms on the surface (see Supporting Information \cite{SI}). Accordingly, the $\sqrt{3}a$ periodicity is $\sim 0.71$ nm, consistent with the expected distance between two nearest-neighbor Fe(1) sites occupying “up” positions in adjacent supercells.

To gain microscopic insight into the origin of the experimentally observed contrast variations in the STM topographs and to establish a direct correspondence between the surface morphology and the underlying atomic configurations, we carried out first-principles DFT calculations. After satisfying the structural relaxation criteria in the DFT calculations, we performed self-consistent ground-state charge density calculations. Subsequently, partial charge density calculations were performed by integrating the electronic states within the energy window from $-0.6$~eV to the Fermi energy ($E_{\mathrm{F}}$), which were used to simulate the STM topography. The simulated STM images exhibit bright and dark regions corresponding to areas of high and low electronic density, respectively.


We present the relaxed atomic structures blended with the DFT-generated STM topographies for the UDD and $1a \times 1a$ configurations in Figures \ref{Figure4}a and \ref{Figure4}b, respectively. The simulated STM images show good agreement with the experimental STM topography displayed in Figures \ref{Figure2}c and \ref{Figure2}e, respectively. The simulated spatial distribution of the Kohn--Sham states near the Fermi energy for the $1a \times 1a$ structure exhibits an approximately circular profile, reflecting the localized character of these states, which predominantly originate from the Te $p_z$ orbitals. In contrast, the corresponding energy-resolved local density of states for the UDD structure is significantly distorted and displays a pronounced dumbbell-like spatial profile. This anisotropic distribution signals enhanced hybridization between the surface Te atoms and their nearest-neighbor Fe atoms in the subsurface layer. Such Fe--Te hybridization modifies the spatial character of the electronic states near the Fermi level, giving rise to the distinct spatial patterns observed in the UDD configuration.

Importantly, these ordering motifs are not merely structural variations; they are expected to significantly influence the electronic band structure and determine the symmetry-permitted magnetic interactions in Fe\textsubscript{5–x}GeTe\textsubscript{2}. Our investigations reveal a strong correlation between the local structural ordering in Fe\textsubscript{5–x}GeTe\textsubscript{2} and its electronic characteristics. This sample shows metallic nature in general, as reported previously \cite{ly2021direct}. However, in this study, a noticeable suppression in the density of states (DOS) is observed near the Fermi energy for the undistorted hexagonal region with Fe vacancy, suggesting the emergence of a small energy gap. Scanning tunneling spectroscopy (STS) measurements performed on regions with two distinct structural orderings, as depicted in Figure \ref{Figure5}(a-b) the region exhibiting the $\sqrt{3}a \times \sqrt{3}a$ superstructure, shows no gap, indicating that the metallic nature is largely preserved for this kind of structural ordering. In contrast, areas corresponding to the undistorted $1a \times 1a$ phase display a significantly prominent gap of about 193 meV, reflecting a pronounced reduction in metallicity, portraying relatively less metallic behaviour. For quantitative estimation, the energy range where the normalized dI/dU signal is almost equal to zero, was considered as the gap region. These findings demonstrate that local structural modulations play a crucial role in governing the low-energy electronic behavior of Fe\textsubscript{5–x}GeTe\textsubscript{2}.

We calculated the electronic band structures for these two different structural configurations of Fe\textsubscript{5-x}GeTe\textsubscript{2}, to understand the experimental results of the systems. Figures \ref{Figure6} and \ref{Figure7} represent the band structures of the $1a \times 1a$ and UDD configurations, respectively. In STM measurements, the tunneling current is governed by the matrix element between the electronic wave functions of the sample and the STM tip, as well as by the local density of states (LDOS) of the sample atoms in the vicinity of the tip, therefore, we focus on the LDOS and tunneling matrix between the tip of the STM and the atoms below the tip.

{All our STM simulations are based on the Tersoff–Hamann approximation~\cite{Tersoff1985}, assuming an s-wave STM tip, and the density of states of the tip is assumed to be energy independent. Consequently, the tunneling current is proportional to the local density of states (LDOS) of the sample evaluated at the tip position near the Fermi energy. Electronic states with finite in-plane crystal momentum $k_{\parallel}$ decay exponentially into the vacuum, with a decay rate that increases with $k_{\parallel}$. As a result, states with small in-plane momentum, i.e., those located near the $\Gamma$ point, decay more slowly into the vacuum and therefore provide the dominant contribution to the tunneling current.

Motivated by this consideration, we focus on electronic states in the vicinity of the $\Gamma$ point to interpret the experimental $dI/dU$ spectra. To account for the finite spatial extent of the STM tip, as well as additional broadening effects, we integrate the electronic contributions within a window corresponding to $7\%$ (which matches the experimental result best) of the total Brillouin zone centered around the $\Gamma$ point.

From the band-structure plots shown in Figure \ref{Figure6} for the $1a \times 1a$ configuration, we clearly observe a gap-like suppression of electronic states around the $\Gamma$ point just above the Fermi energy. In contrast, the band structures presented in Figure \ref{Figure7} for the UDD configuration do not exhibit such a pronounced suppression in the vicinity of $\Gamma$. This qualitative distinction is consistently manifested in the experimental $dI/dU$ spectra shown in Figure~\ref{Figure5}(b), where a suppression of spectral weight 
near zero bias is observed only for the $1a \times 1a$ configuration.

To further elucidate the origin of this behavior, we analyze how the local density of states (LDOS) evolves and how the presence of surface atoms influences the electronic charge density at the surface, which is the key quantity governing the tunneling current in STM measurements. To this end, we plot atom-resolved projected band structures for the atoms located in the topmost layer with respect to the STM tip for the $1a \times 1a$ configuration, as shown in Figure \ref{Figure6}. Three representative atoms, indicated by colored circles in Figure \ref{Figure6}d, are selected. For each of these atoms, we present the corresponding orbital-resolved band structure and the projected density of states (PDOS), obtained by integrating contributions from the shaded region around the $\Gamma$ line.

The projected band structure associated with the top Te atom [Figure \ref{Figure6}a] reveals an absence of electronic states just above the Fermi energy, consistent with the gap-like suppression observed near $\Gamma$. Localized Te $p_z$ states are present on both sides of this gap, while states at higher energies show an increasing contribution from the in-plane $p_x$ and $p_y$ orbitals. Since $p_x$ and $p_y$ orbitals are oriented parallel to the surface, their wave functions decay more rapidly into the vacuum and therefore typically contribute weakly to the STM tunneling current. Nevertheless, a finite experimental $dI/dU$ signal is observed both below and above the Fermi energy even in energy regions where the Te $p_z$ contribution is suppressed.

\begin{figure}[t]
   \centering
       \includegraphics[width=\columnwidth]{"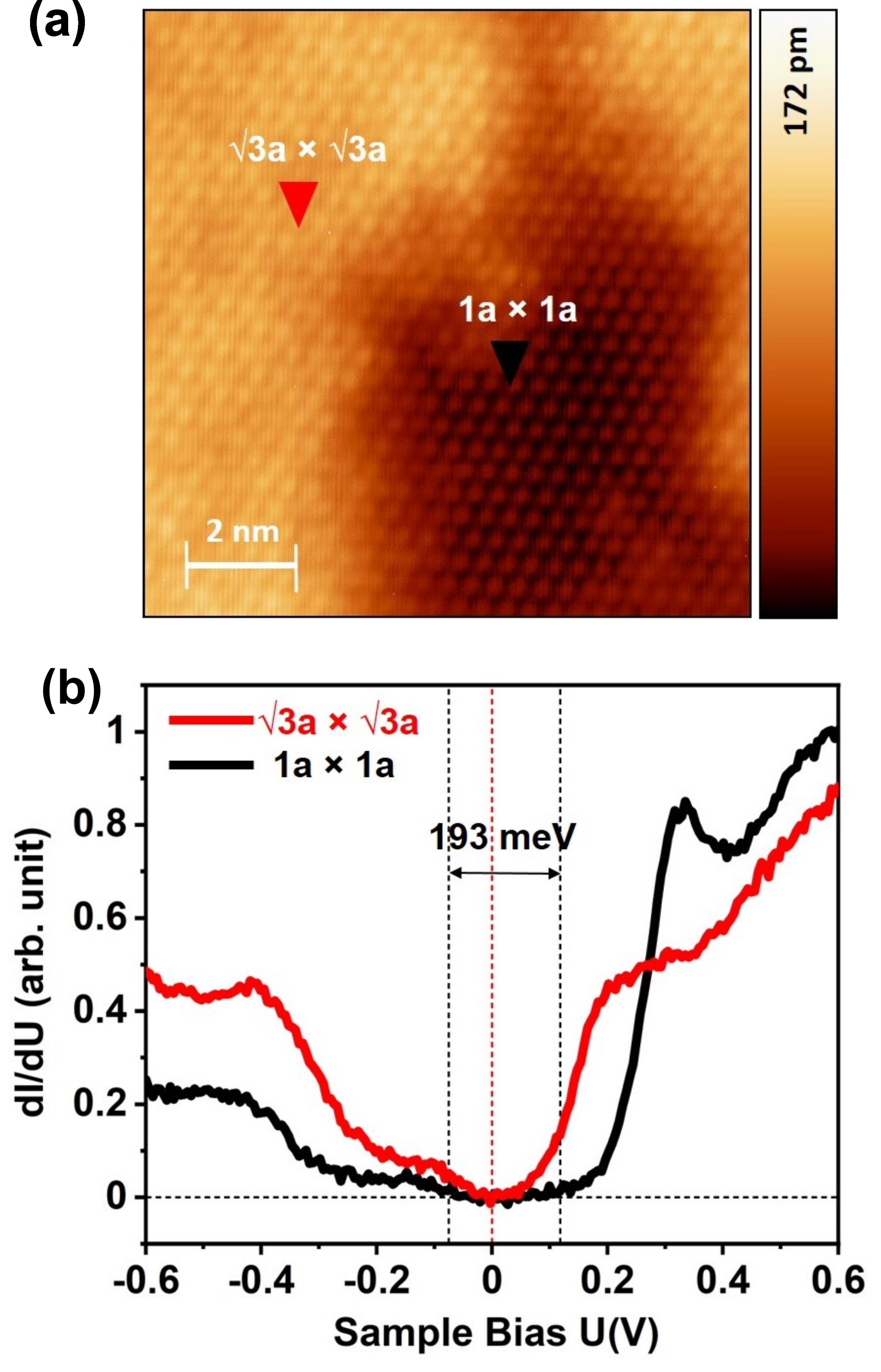"}
    \caption{(a) STM topograph (10 nm × 10 nm) of a region with coexisting $\sqrt{3}a \times \sqrt{3}a$ ordering and the undistorted $1a \times 1a$ phase (U = -200 mV, I = 190 pA). (b) Representative differential conductance spectra acquired on the $\sqrt{3}a \times \sqrt{3}a$ (red) and $1a \times 1a$ (black) regions; the spectroscopic measurement points are indicated in Figure \ref{Figure5}a by red and black triangles, respectively.}
    \label{Figure5}
\end{figure}

This apparent discrepancy can be understood from the orbital-resolved band structures of the Fe atoms shown in Figures \ref{Figure6}(b) and \ref{Figure6}(c), which clearly reveal strong hybridization between Fe $d$ orbitals and Te $p$ orbitals. Such hybridization is symmetry allowed by the $C_{3v}$ point-group symmetry of the crystal. As a consequence of this Fe--Te hybridization, the spatial character of the relevant Kohn--Sham states near the Fermi level is significantly modified, leading to a distortion and partial delocalization of the nominally in-plane Te $p_x$ and $p_y$ orbitals. This hybridization-induced redistribution enhances the amplitude of these states in the vacuum region, thereby enabling them to contribute effectively to the tunneling current.

To visualize this effect directly, we plot the real-space distribution of the energy-integrated local density of states (LDOS), obtained by integrating the LDOS from the Fermi energy up to the applied bias voltage. Figure~\ref{Figure6}(e) shows a relatively low LDOS intensity above the top Te atom for a bias voltage of $+0.2$~eV, in contrast to the higher LDOS intensities observed in Figures~\ref{Figure6}(f) and \ref{Figure6}(g) at other bias voltages. This bias-dependent modulation of the energy-integrated LDOS is fully consistent with the corresponding features observed in the experimental $dI/dU$ spectra.

\begin{figure*}
   \centering
       \includegraphics[width=\textwidth]{"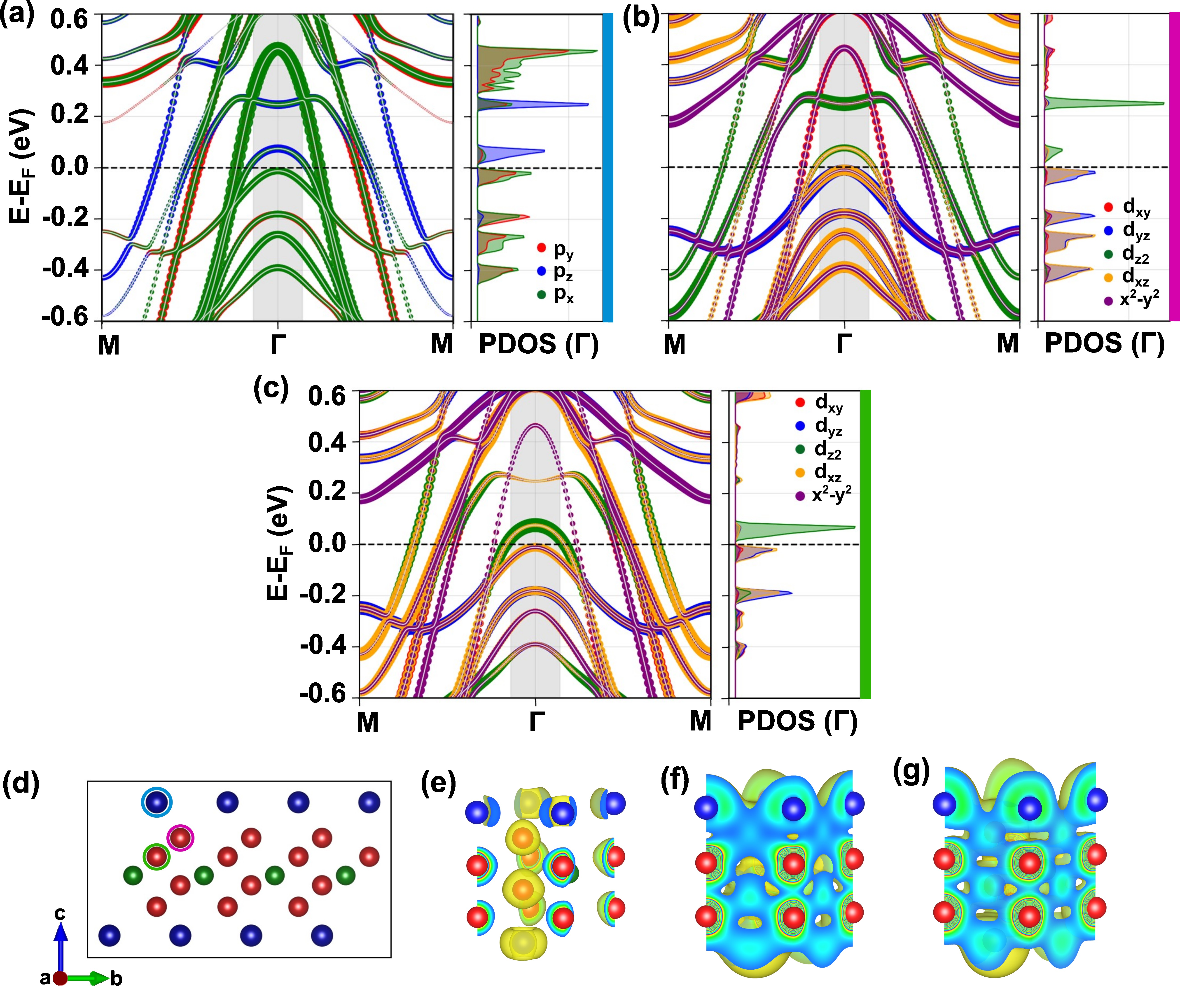"}
    \caption{(a-c) Atom-specific, orbital-resolved band structure plots of the $1a \times 1a$ configuration along the M--$\Gamma$--M direction, including spin--orbit coupling (SOC). The specific atoms and their corresponding band plots are indicated using the same colors, which are also shown as thick lines on the right-hand side of each projected density of states (PDOS) plots, and correspond to the atoms marked by circles in panel (d). Different colors of the bands represent different orbitals within each band plot, and the thickness of each colored band signifies the corresponding orbital weight at a given $\mathbf{k}$ point. The gray rectangle around the $\Gamma$ point in all band plots represents 7\% of the total Brillouin zone, within which the contributions are used to calculate the PDOS shown on the right side of each band plot. The black dashed line denotes the Fermi energy $E_F$. (e--g) Real-space distributions of the energy-integrated local density of states (LDOS) obtained by integrating the LDOS over the energy windows 0 to 0.2~eV, $-0.6$ to 0.0~eV, and 0.0 to 0.6~eV, respectively. All panels are visualized using the same isosurface value of 0.001~arb.~units.}
    \label{Figure6}
\end{figure*}

Similarly, for the UDD configuration, Figure \ref{Figure7}a shows the atom-projected band structure for the top Te atom, while Figures \ref{Figure7}(b-d) present the orbital-resolved band structures for the Fe atoms indicated by different colored circles in Figure \ref{Figure7}e. In contrast to the $1a \times 1a$ configuration, these band structures do not exhibit a pronounced suppression of the LDOS near the $\Gamma$ point. Instead, a finite density of states persists close to the Fermi energy.

\begin{figure*}
   \centering
       \includegraphics[width=\textwidth]{"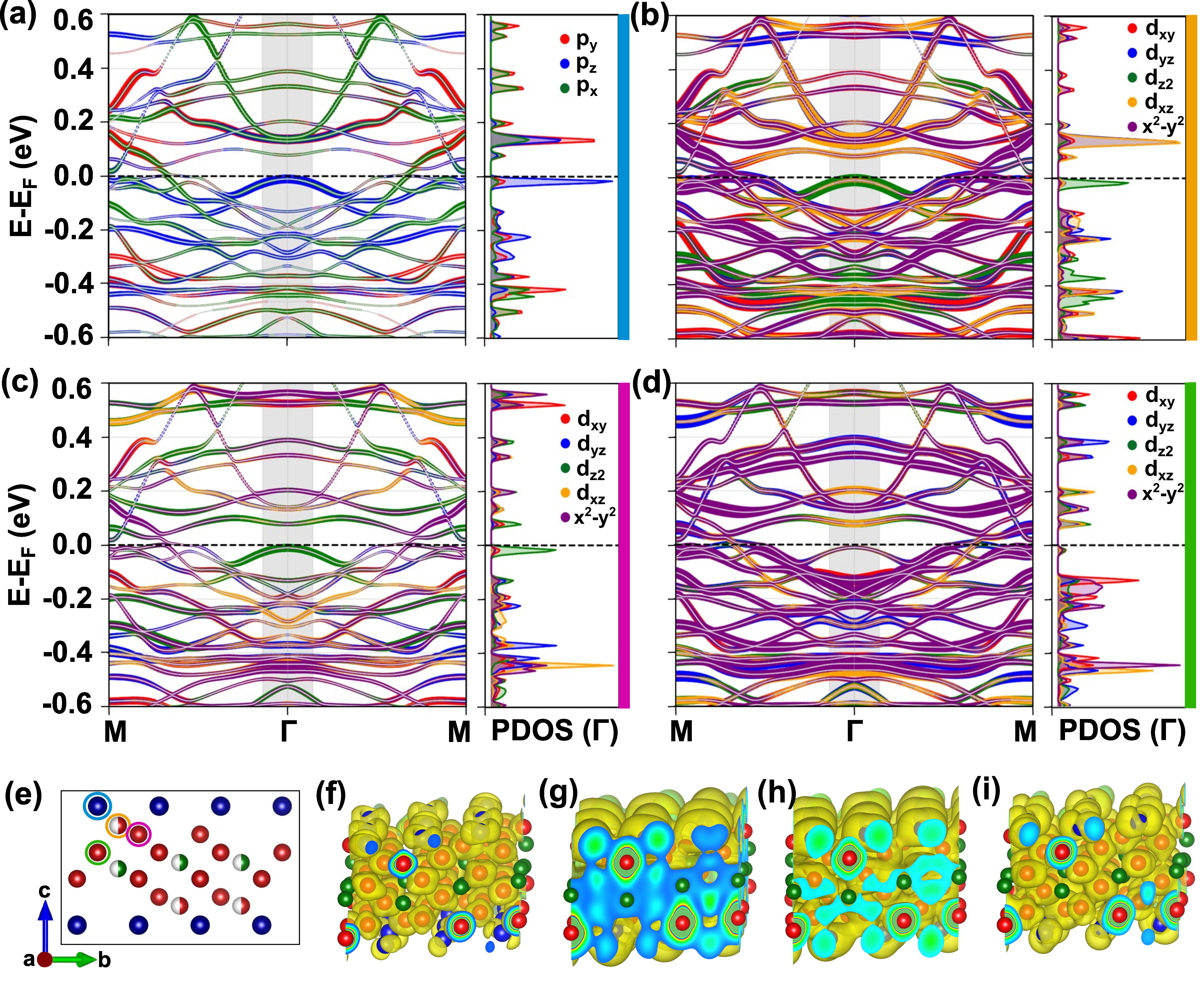"}
       \caption{(a-d) Atom-specific, orbital-resolved band structure plots of the UDD configuration along the M--$\Gamma$--M direction, including spin--orbit coupling (SOC). The specific atoms and their corresponding band plots are indicated using the same colors, which are also shown as thick lines on the right-hand side of each projected density of states (PDOS) plots, and correspond to the atoms marked by circles in panel (e). Different colors of the bands represent different orbitals within each band plot, and the thickness of each colored band signifies the corresponding orbital weight at a given $\mathbf{k}$ point. The gray rectangle around the $\Gamma$ point in all band plots represents 7\% of the total Brillouin zone, within which the contributions are used to calculate the PDOS shown on the right side of each band plot. The black dashed line denotes the Fermi energy $E_F$. (f--i) Real-space distributions of the energy-integrated local density of states (LDOS) obtained by integrating the LDOS over the energy windows 0 to 0.2~eV, 0 to 0.6~eV, $-0.6$ to 0~eV, and $-0.2$ to 0~eV, respectively. All panels are visualized using the same isosurface value of 0.001~arb.~units.}
    \label{Figure7}
\end{figure*}

A key difference in the UDD configuration is the enhanced hybridization involving the Fe-up atom, which is absent or significantly weaker in the $1a \times 1a$ structure. The presence of this Fe atom leads to enhanced hybridization between Fe $d$ orbitals and Te $p$ orbitals, in particular the in-plane $p_x$ and $p_y$ states. Although the Te $p_x$ and $p_y$ orbitals are nominally confined to the surface plane and therefore contribute weakly to STM tunneling due to their rapid decay into the vacuum, their hybridization with Fe $d$ states significantly modifies the spatial character of the corresponding Kohn--Sham states. This hybridization induces a finite out-of-plane component, increasing the vacuum amplitude of these states. As a result, the hybridized Te $p_x$ and $p_y$ states contribute more to the STM $dI/dU$ signal.

\begin{figure*}
   \centering
       \includegraphics[width=\textwidth]{"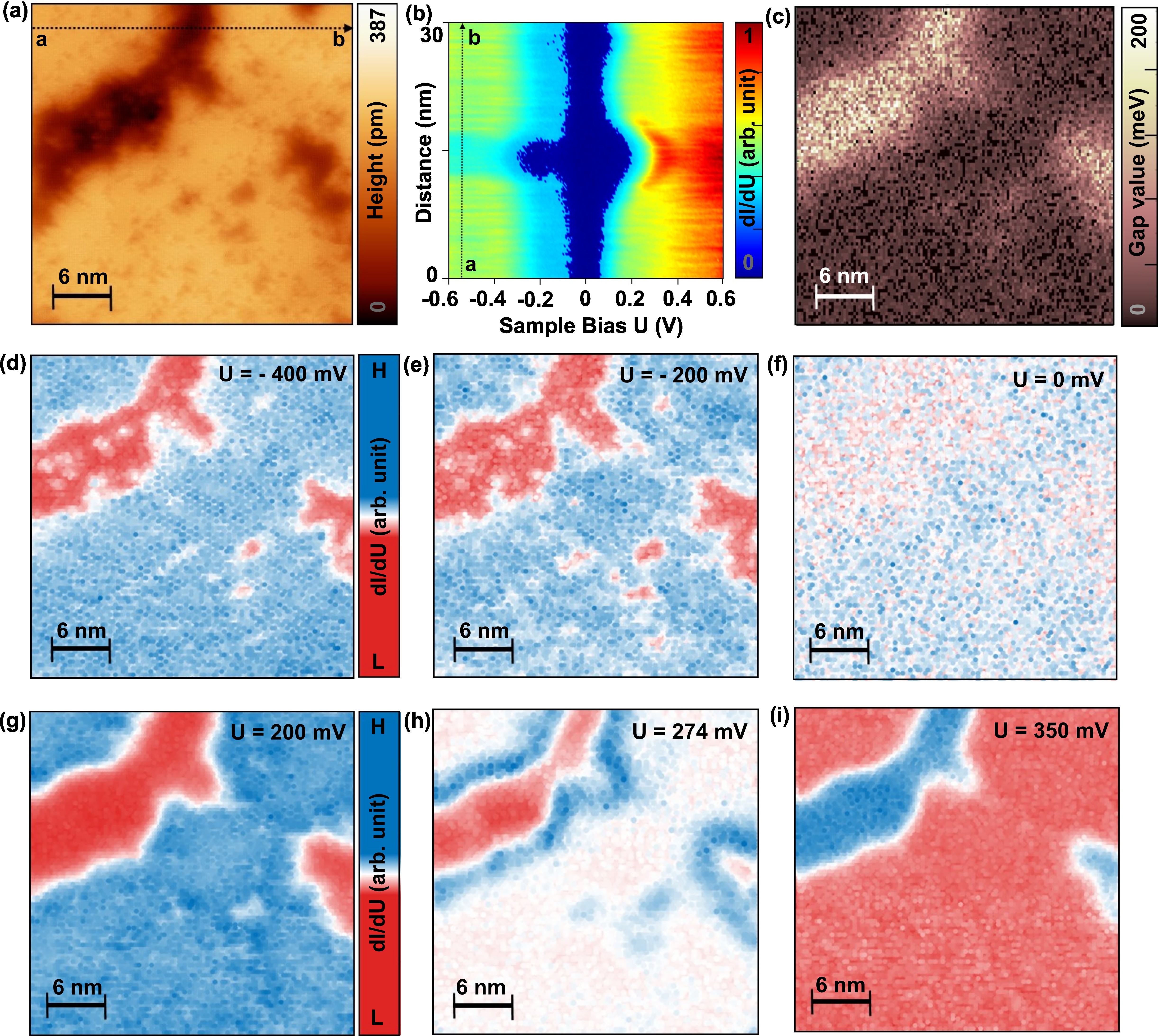"}
    \caption{(a) STM topograph (30 nm × 30 nm) displaying regions (U = -250 mV, I = 50 pA) with $\sqrt{3}a \times \sqrt{3}a$ ordering (bright contrast) and the $1a \times 1a$ phase (dark contrast). (b) 2D projection of a series of 128 STS spectra, acquired along a 30 nm line, marked by the black arrow from point a to b over the surface shown in panel (a), spanning both the structural phases. The color scale denotes the local density of states (LDOS). A pronounced suppression of DOS is observed in the $1a \times 1a$ region compared to $\sqrt{3}a \times \sqrt{3}a$ region as demonstrated by the dark blue region. (c) Map of the spectral gap $\Delta(r)$, obtained over the same region. The color scale indicates the gap values in meV.  (d–i) Differential conductance (dI/dU) maps acquired over the same area at bias voltages corresponding to the transition energies identified in panel 5(b): (d) –400 mV, (e) –200 mV, (f) 0 mV, (g) 200 mV, (h) 274 mV and (i) 350 mV. The maps reveal the evolution of the local density of states across both structural phases as a function of energy.}
    \label{Figure8}
\end{figure*}

We further visualize this effect by plotting the real-space distribution of the energy-integrated local density of states (LDOS) at a fixed isosurface value, illustrating how the applied bias voltage modifies the spatial character of the electronic states near the surface. Figures \ref{Figure7}(f) and \ref{Figure7}(i) show a relatively low LDOS intensity above the top Te atom for bias voltages of $+0.2$~eV and $-0.2$~eV, respectively. In contrast, Figures \ref{Figure7}(g) and \ref{Figure7}(h), corresponding to bias voltages of $+0.6$~eV and $-0.6$~eV, exhibit a significantly enhanced LDOS intensity above the surface.

Although a suppression of the energy-integrated local density of states above the top Te atom is still observed at low bias in the $1a \times 1a$  configuration, its magnitude is considerably weaker than that found in the UDD configuration. This reduced suppression is consistent with the enhanced Fe--Te hybridization in the UDD structure, which modifies the spatial character of the Kohn--Sham states near the Fermi level and sustains a finite tunneling contribution even at low bias.

From these results, we conclude that the STM data, which exhibit a suppression of the LDOS for the $1a \times 1a$ structure—indicative of less metallic behavior, and a comparatively more conductive response for the UDD configuration, are well captured by our density functional theory calculations within the Tersoff--Hamann approximation~\cite{Tersoff1985}. Our analysis further highlights the crucial role of the Fe(1) atom in enhancing hybridization with the top-layer Te atoms, which significantly modifies the Te $p$-orbital character and governs the LDOS probed in the STM measurements.

This contrast is further confirmed in line-scan STS measurements: along a 30 nm trajectory crossing both phases (Figure \ref{Figure8}a, from a to b point along the black arrow), a distinct spatial variation in the DOS is observed, with the more metallic $\sqrt{3}a \times \sqrt{3}a$ regions exhibiting no gap at $E_{\mathrm{F}}$, whereas the $1a \times 1a$ regions show a well-defined gap. This variation is clearly illustrated in Figure \ref{Figure8}b, where the dark blue region denotes the energy gap. A systematic increase in the gap magnitude is evident as the spectra progress from the brighter (more metallic) to the darker (less metallic) regions, confirming its direct dependence on the underlying structural modulation. To quantify the local electronic structure, we acquire $dI/dU$ spectra on a densely spaced pixel grid and create a map \cite{zhao2019atomic} of the approximate gap magnitude $\Delta(r)$ as displayed in Figure \ref{Figure8}c. In agreement with Figure \ref{Figure5}b, the extracted gap magnitude is nearly zero in the $\sqrt{3}a \times \sqrt{3}a$ regions, while the $1a \times 1a$ region displays a gap value close to 200 meV.

To get a spatial variation of the local density of states on the regions with mixed phases, differential conductance maps are measured as a function of energy \cite{sinha2025situ} as described in the Supporting Information \cite{SI}. Figures \ref{Figure8}(d-i) illustrate dI/dU maps obtained at different sample biases, corresponding to the energies at which abrupt changes appear in the DOS curves in Figure \ref{Figure5}b. The two regions are well distinguished in these spectroscopic imaging at all other energies except at Fermi energy. In the valence band region the more metallic phase show higher density of state compared to the less-metallic phase. Similar results is observed in conduction band region upto 300 mV. At 300 mV a complete inversion of LDOS is observed, and at energies $\geq$ 300 mV the insulating phase shows higher DOS. This result is consistent with the DOS vs energy plot in Figures \ref{Figure5}b and \ref{Figure8}b. These spatially resolved spectroscopic maps provide direct evidence of electronic phase separation at the nanoscale, where metallic and pseudogapped domains coexist within the same crystal lattice, as previously observed in case of YbB\textsubscript{6} \cite{coe2025nanoscale}. The persistence of this contrast over a wide energy range further supports the strong coupling between local structural order and the underlying electronic states in Fe\textsubscript{5–x}GeTe\textsubscript{2}.

The coexistence of metallic and pseudogapped regions within the same crystal highlights the sensitivity of the electronic ground state to subtle structural variations, emphasizing the critical role of Fe(1) layer configurations in tuning both the electronic and magnetic properties of Fe\textsubscript{5–x}GeTe\textsubscript{2}. Such nanoscale phase coexistence, often associated with strongly correlated systems, points to emergent electronic inhomogeneity and phase competition. Beyond fundamental interest, this structural sensitivity provides a pathway for engineering electronic functionality in van der Waals magnets by controlling Fe(1) distributions during synthesis or post-growth treatments. This tunability holds significant promise for practical applications, as the coexistence and controllability of metallic and insulating phases are highly relevant for spintronic architectures, phase-change electronic devices, and neuromorphic computing platforms where resistive switching is essential.

\section{Conclusion}

In summary, we uncover a direct correlation between Fe(1) site ordering and the nanoscale electronic landscape of the vdW ferromagnet Fe$_{5-x}$GeTe$_2$ using scanning tunneling microscopy and spectroscopy. Atomic resolved imaging and STM simulation reveals two distinct structural motifs: an undistorted $1a\times1a$ lattice associated with Fe(1) vacancies and a $\sqrt{3}a\times\sqrt{3}a$ ordered phase arising from Fe(1) site ordering. Spatial resolved differential conductance maps demonstrate that these structural motifs host different electronic responses, leading to nanoscale electronic phase separation. While the $\sqrt{3}a \times \sqrt{3}a$ domains retain metallic electronic character, the $1a \times 1a$ regions exhibit a finite energy gap of approximately 193 meV accompanied by a pronounced suppression of local density of states near the Fermi level. The persistence of these spectroscopic signatures over extended spatial regions and across a broad energy range establishes the robustness of this electronic phase separation.
Complementary density functional theory calculations provide microscopic insight into the origin of this behavior. By explicitly incorporating different Fe(1) site configurations, our calculations reproduce the contrasting low-energy electronic responses of the two structural phases, capturing the suppression of the density of states near the Fermi level in the $1a \times 1a$ configuration and the more metallic character of the $\sqrt{3}a \times \sqrt{3}a$ phase. The analysis identifies enhanced Fe–Te orbital hybridization, particularly involving the Fe(up) site, as the key mechanism governing the redistribution of Te p-derived states and the resulting spatial variation in the tunneling spectra.
Our findings highlight the critical role of intrinsic structural disorder in shaping the electronic landscape of vdW magnets and establish Fe$_{5-x}$GeTe$_2$ as a model system for exploring structure-driven electronic inhomogeneity in layered magnetic materials.

\section{Acknowledgement}

S.S. and S.M. sincerely acknowledge IIT Delhi Central Research Facility (CRF) for providing access to scanning tunneling microscope (STM). S.S. acknowledges the Department of Science and Technology (DST), Govt. of India, for providing INSPIRE fellowship [IF190537]. S.S. also thanks Ms. Neelam, Project Scientist, CRF for her assistance with the PPMS magnetic measurements. This work is supported by DST funded project ‘CONCEPT’ under nanomission program (DST/NM/QM-10/2019). RPS acknowledge ANRF funded Core Research Grant (No. CRG/2023/000817). A.J. acknowledge the financial support from DST-India. A.J. and M.K. acknowledge National Supercomputing Mission (NSM) for providing computing resources of ‘PARAM RUDRA’ at S.N. Bose National Centre for Basic Sciences, which is implemented by C-DAC and supported by the Ministry of Electronics and Information Technology (MeitY) and Department of Science and Technology (DST), Government of India.

\bibliographystyle{apsrev4-2-with-title}
\bibliography{references}

\end{document}